\documentclass[twocolumn,aps,prc,superscriptaddress,showpacs,nobibnotes]{revtex4}
\usepackage{epsfig,dcolumn,bm}

\newcommand{\dmu}{\partial_\mu}

\newcommand{\lc}{{\cal L}}

\newcommand{\jc}{{\cal J}}

\newcommand{\sqd}{\sqrt{2}}

\newcommand{\sqs}{\sqrt{6}}

\newcommand{\be}{\begin{equation}}
\newcommand{\ee}{\end{equation}}
\newcommand{\bea}{\begin{eqnarray}}
\newcommand{\eea}{\end{eqnarray}}
\newcommand{\nn}{\nonumber}

\begin{document}

\title{Dynamically generated open and hidden charm mesons}

\author{D. Gamermann}
\affiliation{Departamento de F\'isica Te\'orica and IFIC, Centro Mixto
Universidad de Valencia-CSIC,\\ Institutos de Investigaci\'on de
Paterna, Aptdo. 22085, 46071, Valencia, Spain} 
\author{E. Oset}
\affiliation{Departamento de F\'isica Te\'orica and IFIC, Centro Mixto
Universidad de Valencia-CSIC,\\ Institutos de Investigaci\'on de
Paterna, Aptdo. 22085, 46071, Valencia, Spain} 
\author{D. Strottman}
\affiliation{Departamento de F\'isica Te\'orica and IFIC, Centro Mixto
Universidad de Valencia-CSIC,\\ Institutos de Investigaci\'on de
Paterna, Aptdo. 22085, 46071, Valencia, Spain} 
\author{M. J. Vicente Vacas}
\affiliation{Departamento de F\'isica Te\'orica and IFIC, Centro Mixto
Universidad de Valencia-CSIC,\\ Institutos de Investigaci\'on de
Paterna, Aptdo. 22085, 46071, Valencia, Spain}

\begin{abstract}
In this presentation I explain our framework for dynamically generating resonances from the meson meson interaction. Our model generates many poles in the T-matrix which are associated with known states, while at the same time new states are predicted.
\end{abstract}

\maketitle

\section{Introduction}

The past Recent years have been very exciting for charm spectroscopy, since many experiments are investigating the energy region for open and hidden charm production. As a result many new mesonic states have been found and confirmed by several experiments. \cite{exp1,exp2,exp3,exp4,exp5}.

The fact that some of the newly found states do not fit quark model calculations \cite{qmodel} has opened the discussion about the structure of such states. Some authors suggest a tetraquark interpretation \cite{4quark}, some a mixture between conventional mesons and tetraquarks \cite{mix}, while some others suggest a molecular interpretation \cite{molecular}.

Chiral Lagrangians in unitarized coupled channels has already been efficiently used as a starting point to dynamical generate states \cite{chi1,chi2} where the simple $q\bar q$ picture has failed. In these cases the states are generated dynamically through the interaction and appear as poles in the complex scattering T-matrix. 

In this work we are going to present results coming from the use of a phenomenological Lagrangian \cite{meu1,meu2}, describing the interaction of pseudoscalar and vector mesons in coupled channels. The Lagrangian is based in the $SU(4)$ flavor symmetry, but since $SU(4)$ is not a good symmetry of nature, it will be explicitly broken by taking into account the heavy mass of the exchanged mesons in the underling dynamics of the interaction. In the next section the Lagrangians used are presented and the mathematical framework is briefly explained. Section III presents the most important results and some discussion and conclusions are done in section IV.

%%%%%%%%%%%%%%%%%%%%%%%%%%%%%%%%

\section{Framework}

The pseudoscalar and the vector mesons are described by a 15-plet representation of $SU(4)$. First two fields are constructed with the help of the $SU(4)$ generators, one field for the pseudoscalars and another one for the vector mesons:

\bea
\Phi&=&\sum_{i=1}^{15}{\varphi_i \over \sqd}\lambda_i  \\
\cal{V}_\mu&=&\sum_{i=1}^{15}{v_{\mu i} \over \sqd}\lambda_i 
\eea

Each one of these fields is a 4x4 matrix. The meson assignment for each element of the matrix $\Phi$ is:

\begin{widetext}
\begin{eqnarray}
\Phi =\left( \begin{array}{cccc}
 {\pi^0 \over \sqd}+{\eta \over \sqs}+{\eta_c \over \sqrt{12}} & \pi^+ & K^+ & \bar D^0 \\ & & & \\
 \pi^- & {-\pi^0 \over \sqd}+{\eta \over \sqs}+{\eta_c \over \sqrt{12}} & K^0 & D^- \\& & & \\
 K^- & \bar K^0 & {-2\eta \over \sqs}+{\eta_c \over \sqrt{12}} & D_s^- \\& & & \\
 D^0 & D^+ & D_s^+ & {-3\eta_c \over \sqrt{12}} \\ \end{array} \right)   \\
\end{eqnarray}
\end{widetext}
and for the the matrix $\cal{V}_\mu$:
\begin{widetext}
\begin{eqnarray}
V^\mu=\left( \begin{array}{cccc}
{\rho_\mu^0 \over \sqd}+{\omega_\mu \over \sqs}+{J/\psi_{ \mu} \over \sqrt{12}} & \rho^+_\mu & K^{*+}_\mu & \bar D^{*0}_\mu \\ & & & \\
 \rho^{*-}_\mu & {-\rho^0_\mu \over \sqd}+{\omega_\mu \over \sqs}+{J/\psi_{\mu} \over \sqrt{12}} & K^{*0}_\mu & D^{*-}_\mu \\& & & \\
  K^{*-}_\mu & \bar K^{*0}_\mu & {-2\omega_\mu \over \sqs}+{J/\psi_{\mu} \over \sqrt{12}} & D_{s\mu}^{*-} \\& & & \\
D^{*0}_\mu & D^{*+}_\mu & D_{s\mu}^{*+} & {-3J/\psi_\mu \over \sqrt{12}} \\ \end{array} \right).   
\end{eqnarray}
\end{widetext}

Now for each field we define a hadronic current:

\bea
J_\mu&=&(\dmu \Phi)\Phi-\Phi\dmu\Phi \\
\cal{J}_\mu&=&(\dmu \cal{V}_\nu)\cal{V}^\nu-\cal{V}_\nu\dmu \cal{V}^\nu . \label{curj}
\eea

The Lagrangians are built by connecting these currents:

\bea
\lc_{PPPP}={1\over12f^2}Tr({J}_\mu {J}^\mu+\Phi^4 M) \label{lag} \\
\lc_{PPVV}={-1\over 4f^2}Tr\left(J_\mu\cal{J}^\mu\right). \label{lagini}
\eea

In the Lagrangian of eq. (\ref{lag}) we take $M=\textrm{diag}\left(m_\pi^2,m_\pi^2,2m_K^2-m_\pi^2,2m_D^2-m_\pi^2\right)$, in this way we reproduce the lowest order chiral Lagrangian in the light sector.

One can interpret the interaction vertices in the Lagrangians (\ref{lag},\ref{lagini}) as exchanges of vector mesons. Hadronic currents not conserving charm (carrying explicit charm quantum number) should be connected by the exchange of heavy mesons, since in this case the exchanged particle must contain a charmed quark. In some sectors of the hidden charm interaction it is possible to exchange also a hidden charm meson, which is also heavy. In this cases it is possible to separate the contribution of the heavy and light exchanged particle. Once we have identified the exchanges of heavy particles we suppress such terms, because of the heavy mass of these particles. For suppressing these terms we use the square of the ratio between the light and heavy vector meson mass, as it would occur in the propagator of the exchanged vector meson. The final Lagrangians read:

\begin{widetext}
\begin{eqnarray}
\lc_{PPPP}&=&{1\over 12 f^2}\biggr(Tr\Big(J_{88\mu}J_{88}^\mu+2 J_{3 \bar 3\mu}J_{88}^\mu+J_{3 \bar 3_\mu} J_{3 \bar 3}^\mu\Big)+{8\over3}\gamma J_{\bar 3 1\mu}J_{1 3}^\mu\nonumber\\
&+&{4\over\sqrt{3}}\gamma\Big(J_{\bar 3 1\mu}J_{8 3}^\mu+J_{\bar 3 8\mu}J_{1 3}^\mu\Big)+2\gamma J_{\bar 3 8\mu}J_{8 3}^\mu+\psi J_{\bar 3 3_\mu} J_{\bar 3 3}^\mu+Tr\left(M \Phi^4 \right)\biggr)  \label{lagfully}. \\
\lc_{PPVV}&=&{-1\over 4f^2}\Big( Tr\bigg(J_{88\mu}\jc_{88}^\mu+J_{3\bar3\mu}\jc_{3\bar3}^\mu+J_{88_\mu}\jc_{3\bar3}^\mu+
 J_{3\bar3\mu}\jc_{88}^\mu+\gamma J_{83\mu}\jc_{\bar3 8}^\mu \nonumber \\
& +&{2\gamma\over\sqrt{3}}(J_{83\mu}\jc_{\bar3 1}^\mu+ J_{13\mu}\jc_{\bar38}^\mu)+
{4\gamma\over3}J_{13\mu}\jc_{\bar3 1}^\mu\bigg)+\psi J_{\bar3 3\mu}\jc_{\bar3 3}^\mu \nonumber \\
 &+ &\gamma J_{\bar3 8\mu}\jc_{83}^\mu+{2\gamma\over\sqrt{3}}(J_{\bar3 8\mu}\jc_{13}^\mu+J_{\bar3 1\mu}\jc_{83}^\mu)+
 {4\gamma\over3}J_{\bar3 1\mu}\jc_{13}^\mu \Big), \label{lag2}
\end{eqnarray}
\end{widetext}
where
\begin{eqnarray}
J_{ij\mu}&=&(\dmu\phi_i)\phi_j-\phi_i\dmu\phi_j \\
\jc_{ij\mu}&=&(\dmu V_{i\nu})V_j^\nu-V_{i\nu}\dmu V_j^\nu \\
\gamma&=&\left( {m_L\over m_H}\right)^2 \\
\psi&=&-{1\over3}+{4\over3} \left( {m_L\over m_{J/\psi}}\right)^2
\end{eqnarray}
and $\phi_i$ and $V_i^\mu$ with $i=1,3,\bar3,8$ are the $SU(3)$ fields into which the 15-plets of $SU(4)$ break, the exact meson assignment for each one of these fields can be found in \cite{meu1,meu2}.

With these Lagrangians, applying usual Feynman rules, one can get tree level transition amplitudes between any possible two meson initial and final states. All possible two particle states from the two 15-plets with the same quantum numbers form a coupled channel space.

The amplitudes are projected in s-wave and collected in a matrix for all channels in each coupled channel space. This matrix, that we call potential, is plugged as the kernel for solving the Bethe-Saltpeter (BS) equation that in an on-shell formalism assumes an algebraic form \cite{oller}.

The scattering of pseudoscalars out of pseudoscalars in s-wave has the quantum numbers of a scalar. The BS-equation in this case has the solution:

\begin{equation}
T=(\hat 1-VG)^{-1}V.
\end{equation}
In this equation $V$ is the potential. The matrix G is diagonal with each one of its non-zero elements given by the loop function for the two particles in each channel:

\begin{widetext}
\begin{eqnarray} 
 G_{ii}&=&{1 \over 16\pi ^2}\biggr( \alpha _i+Log{m_1^2 \over \mu ^2}+{m_2^2-m_1^2+s\over 2s}
  Log{m_2^2 \over m_1^2}
 + {p\over \sqrt{s}}\Big( Log{s-m_2^2+m_1^2+2p\sqrt{s} \over -s+m_2^2-m_1^2+
  2p\sqrt{s}}\nn\\
&+&Log{s+m_2^2-m_1^2+2p\sqrt{s} \over -s-m_2^2+m_1^2+  2p\sqrt{s}}\Big)\biggr)
  \label{loopf}
\end{eqnarray}
\end{widetext}
In equation (\ref{loopf}) $m_1$ and $m_2$ are the masses of the two mesons in channel $i$.
Over the real axis $p$ is the three-momentum of the mesons in the center of mass frame:

\begin{eqnarray}
p&=&{\sqrt{(s-(m_1+m_2)^2)(s-(m_1-m_2)^2)}\over 2\sqrt{s}} \label{trimom}.
\end{eqnarray}
In the complex plane the momentum $p$ is calculated using the same expression.

For the interaction of the pseudoscalars with the vector mesons, one has to take into account the polarization of the vector mesons. In this case the unitarized T-matrix assumes the form \cite{chi1,chi2}:

\begin{eqnarray}
 T&=&-({\hat 1} + V{\hat G})^{-1}V \overrightarrow{\epsilon}.\overrightarrow{\epsilon}' \label{bseq2}
\end{eqnarray}
In this equation $\hat G$ is a diagonal matrix with each element given by:

\begin{eqnarray}
\hat G_{ii}&=&G_{ii}\left(1+{p^2\over3m_2^2}\right) \label{loopvec}
\end{eqnarray}
here $G_{ii}$ is the usual loop function given by equation (\ref{loopf}) with $m_1$ the mass of the pseudoscalar meson and $m_2$ the mass of the vector meson.

The loop function has two non-independent parameters, $\alpha_i$ and $\mu$. Since they are not independent we will fix one, $\mu$=1500 MeV and fit only the subtraction constants, $\alpha_i$, as free parameters of the theory.

The loop function has, over the real axis, the right imaginary part to ensure the unitarity of the T-matrix \cite{chi1}:

\begin{equation}
Im(G_{ii})=-{p\over 8\pi\sqrt{s}} \label{imloop}.
\end{equation}

With this framework one is able to solve the BS-equation and study the T-matrix in order to identify poles in the second Riemann sheet. The poles identified can be associated with know resonances, as we describe in the following section.

\section{results}

At first we studied the $SU(3)$ structure of the interaction. The 15-plet of $SU(4)$ breaks down into four multiplets of the lower symmetry: an octet and a singlet with null charm quantum number and an antitriplet and a triplet with positive and negative charm, respectively. Apart from the octets and singlets with null charm coming from the scattering of two light octets, one can also expect hidden charm singlets coming from the scattering of the heavy triplets with the antitriplets. This results in one hidden charm scalar resonance and two axial ones. In the open charm sector one can expect to generate poles corresponding to antitriplets and sextets.

For the scalars, in the open charm sector, the antitriplet generated has two isospin members that can be identified with the $D_{s0}^*(2317)$ and $D_0^*(2400)$. The sextet generated comes out too broad, so one cannot expect to be able to identify it experimentally. We should notice however, that in other works \cite{lutz1,chiang1} this sextet appears as narrow states. With respect to these other works we take into account the different meson decay constant for the pion and the $D$ mesons, plus the exchange of heavy vector mesons. In \cite{meu1} we have also used a different Lagrangian with different pattern of $SU(4)$ symmetry breaking borrowed from \cite{skyrme}. The differences obtained with the two models indicate the stability of the results. The $D_{s0}^*(2317)$ and $D_0^*(2400)$ of the antitriplet are rather stable, but there are some larger differences in the states of the sextet, which in any case appear always with a very broad width.

Still in the scalar sector one can observe a hidden charm resonance coming mainly from the interaction between the $D$ and $\bar D$ mesons. This resonance appears around 3.7 GeV and is very narrow. Our model also reproduces the light octet and singlet of scalar mesons to be identified with the $f_0(980)$, $a_0(980)$, $\kappa$ and $\sigma$.

The spectrum for the axial resonances is richer, since in this case one can distinguish between a light pseudoscalar scattering with a heavy vector and a light vector scattering with a heavy pseudoscalar. So in the open charm sector two antitriplets are generated, one of them was identified with the $D_{s1}(2460)$ and $D_1(2430)$ while the other one with the $D_{s1}(2536)$ and $D_1(2420)$ resonances. Furthermore in the present case, one of the two sextets generated in this sector is very broad but the second one has relatively smaller width ($\sim$200 MeV), so in principle these exotic states could be observed.

In the light axial sector two octets and a singlet are generated. One of the octets is associated with the $K_1(1270)$, $b_1(1235)$ and $h_1(1380)$ states while the second one we associate with $K_1(1270)$, $a_1(1260)$ and $f_1(1285)$ states. Note that this association considers that the $K_1(1270)$ has a two pole structure \cite{geng}. The light singlet is identified as the $h_1(1170)$.

For the hidden charm sector two singlets can be identified, they are states of defined G-parity. The positive state is identified with the $X(3872)$ state, while the negative G-parity state is a prediction of our model. In \cite{meu2} the subtraction constant in the loop function was set to -1.55, in this case the two resonances appear around 3.84 GeV and are degenerated by 3 MeV. If one takes a bigger subtraction constant, around -1.3, the two states are nearly degenerated ($M_+$-$M_-<$0.5 MeV) around the experimental value of 3.87 GeV. In this situation only experiments that can select definite G(or C)-parity would be able to differentiate between the two.

\section{Conclusion}

We briefly presented and explained our model for generating dynamically poles in the complex scattering T-matrix. These poles can be associated with many known states, giving therefore a plausible explanation for their structures. Our model starts from a symmetrical $SU(4)$ flavor Lagrangian. Then, using the fact that $SU(4)$ is not a very good symmetry in nature we explicitly break this symmetry down to $SU(3)$ considering that the interaction is driven by exchanges of vector mesons and suppressing terms where the exchanged meson is a heavy one. We also use a different pattern of $SU(4)$ breaking following \cite{skyrme} which allows us to qualify the stability of some results.

In the light sector our model reproduces the results of previous works \cite{chi1,chi2}. In the open charm sector some results coincide with previous works, but the sextets that we generate are broader than in \cite{lutz1,chiang1,lutz2,chiang2}. The spectrum we generate is also richer that in this previous works because we also considered the scattering of light vector mesons with heavy pseudoscalars, which was not calculated by other authors. Our model also allow us to study the hidden charm sector. In this sector we could dynamically generate the $X(3872)$ state and also make prediction of two other states, an axial one with opposite G-parity than that of the $X(3872)$ but which is nearly degenerated in mass, and a scalar resonance with mass around 3.7 GeV.

The picture of dynamically generated states is a very promising one in order to describe many of the newly discovered states. As one can see a very rich and plausible spectrum is generated by this framework taking into account a very simple interaction between the mesons. We hope that the experimental observation of some of the predicted states will enforce this picture.

\end{document}